\newcommand{\avg}[1]{\langle{#1}\rangle}

\newcommand{\req}[1]{(\ref{#1})}

\newcommand{\beq}{\begin{equation}}
\newcommand{\beqar}{\begin{eqnarray}}
\newcommand{\eeqar}{\end{eqnarray}}
\newcommand{\beqars}{\begin{eqnarray*}}
\newcommand{\eeqars}{\end{eqnarray*}}
\newcommand{\eeq}{\end{equation}}
\documentstyle[twoside]{article}

\catcode`\@=11
\long\def\@makefntext#1{
\protect\noindent \hbox to 3.2pt {\hskip-.9pt  
$^{{\eightrm\@thefnmark}}$\hfil}#1\hfill}		

\def\@makefnmark{\hbox to 0pt{$^{\@thefnmark}$\hss}}	
	
\def\ps@myheadings{\let\@mkboth\@gobbletwo
\def\@oddhead{\hbox{}
\rightmark\hfil\eightrm\thepage}   
\def\@oddfoot{}\def\@evenhead{\eightrm\thepage\hfil
\leftmark\hbox{}}\def\@evenfoot{}
\def\sectionmark##1{}\def\subsectionmark##1{}}

\oddsidemargin=\evensidemargin
\newcounter{sectionc}\newcounter{subsectionc}\newcounter{subsubsectionc}
\renewcommand{\section}[1] {\vspace{12pt}\addtocounter{sectionc}{1} 
\setcounter{subsectionc}{0}\setcounter{subsubsectionc}{0}\noindent 
	{\tenbf\thesectionc. #1}\par\vspace{5pt}}
\renewcommand{\subsection}[1] {\vspace{12pt}\addtocounter{subsectionc}{1} 
	\setcounter{subsubsectionc}{0}\noindent 
	{\bf\thesectionc.\thesubsectionc. {\kern1pt \bfit #1}}\par\vspace{5pt}}
\renewcommand{\subsubsection}[1] {\vspace{12pt}\addtocounter{subsubsectionc}{1}
	\noindent{\tenrm\thesectionc.\thesubsectionc.\thesubsubsectionc.
	{\kern1pt \tenit #1}}\par\vspace{5pt}}
\newcommand{\nonumsection}[1] {\vspace{12pt}\noindent{\tenbf #1}
	\par\vspace{5pt}}

\newcounter{appendixc}
\newcounter{subappendixc}[appendixc]
\newcounter{subsubappendixc}[subappendixc]
\renewcommand{\thesubappendixc}{\Alph{appendixc}.\arabic{subappendixc}}
\renewcommand{\thesubsubappendixc}
	{\Alph{appendixc}.\arabic{subappendixc}.\arabic{subsubappendixc}}

\renewcommand{\appendix}[1] {\vspace{12pt}
        \refstepcounter{appendixc}
        \setcounter{figure}{0}
        \setcounter{table}{0}
        \setcounter{lemma}{0}
        \setcounter{theorem}{0}
        \setcounter{corollary}{0}
        \setcounter{definition}{0}
        \setcounter{equation}{0}
        \renewcommand{\thefigure}{\Alph{appendixc}.\arabic{figure}}
        \renewcommand{\thetable}{\Alph{appendixc}.\arabic{table}}
        \renewcommand{\theappendixc}{\Alph{appendixc}}
        \renewcommand{\thelemma}{\Alph{appendixc}.\arabic{lemma}}
        \renewcommand{\thetheorem}{\Alph{appendixc}.\arabic{theorem}}
        \renewcommand{\thedefinition}{\Alph{appendixc}.\arabic{definition}}
        \renewcommand{\thecorollary}{\Alph{appendixc}.\arabic{corollary}}
        \renewcommand{\theequation}{\Alph{appendixc}.\arabic{equation}}
        \noindent{\tenbf Appendix#1}\par\vspace{5pt}}
\newcommand{\subappendix}[1] {\vspace{12pt}
        \refstepcounter{subappendixc}
        \noindent{\bf Appendix \thesubappendixc. {\kern1pt \bfit #1}}
	\par\vspace{5pt}}
\newcommand{\subsubappendix}[1] {\vspace{12pt}
        \refstepcounter{subsubappendixc}
        \noindent{\rm Appendix \thesubsubappendixc. {\kern1pt \tenit #1}}
	\par\vspace{5pt}}

\newcommand{\textlineskip}{\baselineskip=13pt}
\newcommand{\smalllineskip}{\baselineskip=10pt}

\def\eightcirc{
\begin{picture}(0,0)
\put(4.4,1.8){\circle{6.5}}
\end{picture}}
\def\eightcopyright{\eightcirc\kern2.7pt\hbox{\eightrm c}} 

\newcommand{\copyrightheading}[1]
	{\vspace*{-2.5cm}\smalllineskip{\flushleft
	{\footnotesize Mathematical Models and Methods in Applied Sciences #1}\\
	{\footnotesize $\eightcopyright$\, World Scientific Publishing
	 Company}\\
	 }}

\newcommand{\pub}[1]{{\begin{center}\footnotesize\smalllineskip 
	#1\\		
	\end{center}
	}}

\def\abstracts#1#2#3{{
	\centering{\begin{minipage}{4.5in}\baselineskip=10pt\footnotesize
	\parindent=0pt #1\par 
	\parindent=15pt #2\par
	\parindent=15pt #3
	\end{minipage}}\par}} 

\renewenvironment{thebibliography}[1]
	{\frenchspacing
	 \ninerm\baselineskip=11pt
	 \begin{list}{\arabic{enumi}.}
        {\usecounter{enumi}\setlength{\parsep}{0pt}     
	 \setlength{\leftmargin 12.7pt}{\rightmargin 0pt} 
         \setlength{\itemsep}{0pt} \settowidth
	{\labelwidth}{#1.}\sloppy}}{\end{list}}

\newcounter{itemlistc}
\newcounter{romanlistc}
\newcounter{alphlistc}
\newcounter{arabiclistc}

\newcommand{\fcaption}[1]{
        \refstepcounter{figure}
        \setbox\@tempboxa = \hbox{\footnotesize Fig.~\thefigure. #1}
        \ifdim \wd\@tempboxa > 5in
           {\begin{center}
        \parbox{5in}{\footnotesize\smalllineskip Fig.~\thefigure. #1}
            \end{center}}
        \else
             {\begin{center}
             {\footnotesize Fig.~\thefigure. #1}
              \end{center}}
        \fi}

\newcommand{\tcaption}[1]{
        \refstepcounter{table}
        \setbox\@tempboxa = \hbox{\footnotesize Table~\thetable. #1}
        \ifdim \wd\@tempboxa > 5in
           {\begin{center}
        \parbox{5in}{\footnotesize\smalllineskip Table~\thetable. #1}
            \end{center}}
        \else
             {\begin{center}
             {\footnotesize Table~\thetable. #1}
              \end{center}}
        \fi}

\def\@citex[#1]#2{\if@filesw\immediate\write\@auxout
	{\string\citation{#2}}\fi
\def\@citea{}\@cite{\@for\@citeb:=#2\do
	{\@citea\def\@citea{,}\@ifundefined
	{b@\@citeb}{{\bf ?}\@warning
	{Citation `\@citeb' on page \thepage \space undefined}}
	{\csname b@\@citeb\endcsname}}}{#1}}

\newif\if@cghi
\def\cite{\@cghitrue\@ifnextchar [{\@tempswatrue
	\@citex}{\@tempswafalse\@citex[]}}
\def\citelow{\@cghifalse\@ifnextchar [{\@tempswatrue
	\@citex}{\@tempswafalse\@citex[]}}
\def\@cite#1#2{{$\null^{#1}$\if@tempswa\typeout
	{IJCGA warning: optional citation argument 
	ignored: `#2'} \fi}}

\def\pmb#1{\setbox0=\hbox{#1}
	\kern-.025em\copy0\kern-\wd0
	\kern.05em\copy0\kern-\wd0
	\kern-.025em\raise.0433em\box0}

\def\fnt#1#2{\footnotetext{\kern-.3em
	{$^{\mbox{\scriptsize #1}}$}{#2}}}

\def\fpage#1{\begingroup
\voffset=.3in
\thispagestyle{empty}\begin{table}[b]\centerline{\footnotesize #1}
	\end{table}\endgroup}

\def\runninghead#1#2{\pagestyle{myheadings}
\markboth{{\protect\footnotesize\it{\quad #1}}\hfill}
{\hfill{\protect\footnotesize\it{#2\quad}}}}
\font\tenrm=cmr10
\font\tenit=cmti10 
\font\tenbf=cmbx10
\font\bfit=cmbxti10 at 10pt
\font\ninerm=cmr9

\font\eightrm=cmr8

\def\qed{\hbox{${\vcenter{\vbox{			
   \hrule height 0.4pt\hbox{\vrule width 0.4pt height 6pt
   \kern5pt\vrule width 0.4pt}\hrule height 0.4pt}}}$}}

\def\theequation{\thesectionc.\arabic{equation}}	

\begin{document}

\runninghead{Optimal Investment Strategy for Risky Assets}
{Optimal Investment Strategy for Risky Assets}

\normalsize\textlineskip
\thispagestyle{empty}
\setcounter{page}{1}

\copyrightheading{}                     

\vspace*{0.88truein}

\fpage{1}
\centerline{\bf OPTIMAL INVESTMENT STRATEGY FOR RISKY ASSETS}

\vspace*{0.37truein}
\centerline{\footnotesize SERGEI MASLOV}

\vspace*{0.015truein}
\centerline{\footnotesize\it  Department of Physics, 
Brookhaven National Laboratory, Upton, NY, 11973, USA}

\vspace*{10pt}
\centerline{\footnotesize YI-CHENG ZHANG}
\vspace*{0.015truein}
\centerline{\footnotesize\it Institut de Physique Th\'eorique, 
Universit\'e de Fribourg, CH-1700, Switzerland}

\vspace*{0.225truein}
\pub{Received } 

\vspace*{0.21truein}
\abstracts{We design an optimal strategy for investment in a portfolio
of assets subject to a multiplicative Brownian motion. The
strategy provides the maximal typical long-term growth rate of 
investor's capital. We determine the optimal fraction of capital 
that an investor should keep in risky assets as well as 
weights of different assets in an optimal portfolio. In this approach
both average return and volatility of an asset are relevant indicators 
determining its optimal weight. Our results are particularly relevant 
for very risky assets when traditional continuous-time Gaussian 
portfolio theories are no longer applicable. }{}{} 



\vspace*{1pt}\textlineskip

\section{Introduction}
\noindent
The simplest version of the problem  we are going to address in this 
manuscript is rather easy to formulate. Imagine that you are an investor 
with some starting capital, which you can invest in just 
one risky asset. You decided to 
use the following simple strategy: you always maintain a given 
fraction $0<r<1$ of your total current capital invested in this 
asset, while the rest 
(given by the fraction $1-r$) you wisely keep in cash.
You select a unit of time (say a week, a month, a quarter, or a year, 
depending on how closely you follow your investment, and what 
transaction costs are involved) at which you check the asset's current 
price, and sell or buy some shares of this asset. By this transaction
you adjust the current money equivalent of your investment 
to the above pre-selected fraction of your total capital. 

The question we are interested in is: which investment fraction
provides the optimal {\it typical} long-term growth 
rate of investor's capital?  
By typical we mean that this growth rate occurs at large-time 
horizon in majority of realizations of the multiplicative 
process. By extending time-horizon one
can make this rate to occur with probability arbitrary close to one.
Contrary to the traditional economics approach, where the 
expectation value of an artificial ``utility function'' of an investor is 
optimized, the optimization of a typical growth rate does not 
contain any ambiguity.

In this work we also assume that during on the timescale, 
at which the investor checks and readjusts his 
asset's capital to the 
selected investment fraction, the asset's price changes by a random
factor, drawn from some probability distribution, 
and uncorrelated from price dynamics at earlier intervals. 
In other words, the price of an asset experiences a 
{\it multiplicative} random walk with some known probability 
distribution of steps. This assumption is known to hold in real 
financial markets beyond a certain time scale\cite{bouchaud}. 
Contrary to continuum theories popular among economists\cite{merton} our 
approach is not limited to Gaussian distributed returns: indeed, we were 
able to formulate our strategy for a general probability distribution of 
returns per capital (elementary steps of the multiplicative random 
walk).

Our purpose here is to illustrate the essential framework through
simplest examples. Thus risk-free interest rate, asset'sdividends, 
and transaction costs
are ignored (when volatility is large they are indeed negligible).
However, the task of including these effects in our formalism 
is rather straightforward.

The quest of finding a strategy, which optimizes the long-term
growth rate of the capital is by no means new: indeed it was
first discussed by Daniel Bernoulli in about 1730 in connection with
the St. Petersburg game\cite{bernoulli}.  
In the early days of information sciences, 
Shannon\cite{shannon} has 
considered the application of the  concept 
of information entropy in designing optimal strategies in 
such games as gambling. 
Working from the foundations of Shannon,
Kelly has specifically designed an optimal gambling strategy in placing
bets\cite{kelly}, when a gambler has some incomplete information 
about the winning outcome (a ``noisy information channel''). 
In modern day finance, especially the investment in
very risky assets is no different from gambling. The point Shannon and
Kelly wanted to make is that, given that the odds are slightly in your
{\it favor} albeit with large uncertainty, the gambler should not bet 
his whole capital at every time step. On the other hand, 
he would achieve the biggest long-term capital growth by
betting some specially optimized fraction of his whole capital 
in every game. This cautious approach to investment is 
recommended in situations
when the volatility is very large. For instance, in many emergent
markets the volatility is huge, but they are still swarming with
investors, since the long-term return rate in some cautious investment 
strategy is favorable. 

Later on Kelly's approach was expanded and 
generalized in the works of
Breiman\cite{breiman}. Our results for multi-asset optimal investment
are in agreement with his exact but non-constructive equations.
In some special cases, Merton and Samuelson\cite{merton}
have considered the problem of portfolio optimization, when the underlying
asset is subject to a multiplicative {\it continuous} Brownian motion 
with Gaussian price fluctuations. 
Overall, we feel that the topic of optimal long-term investment
has not been adequately exploited, and many interesting consequences 
are yet to be revealed.

The plan of this paper is as follows: in Section 2 we determine the
optimal investment fraction in an (unrealistic) situation when an
investor is allowed to invest in only one risky asset. 
The Section 3 generalizes these results for a more
realistic case when an investor can keep his 
capital in a multi-asset portfolio. In this case
we determine the optimal weights of different assets in this 
portfolio.  

\section{Optimal investment fraction for one asset} 
\noindent
We first consider a situation, when an investor can spend 
a fraction of his capital to buy shares of 
just one risky asset. The rest of his money he keeps in cash.
Generalizing  Kelly\cite{kelly}, we consider the following 
simple strategy of the investor:  he regularly checks the asset's 
current price $p(t)$,  and sells or
buys some asset shares in order to keep the current market value 
of his asset holdings a pre-selected fraction $r$ of his
total capital. These readjustments are
made periodically at a fixed interval, which we refer to as
readjustment interval, and select it as the discrete unit of time.   
In this work the readjustment time interval is selected once and for 
all, and we do not attempt optimization of its length.

We also assume that on the time-scale of this readjustment 
interval the asset price $p(t)$ undergoes a geometric
Brownian motion:
\beq
p(t+1)=e^{\eta (t)} p(t),
\label{price}
\eeq
i.e. at each time step the random number $\eta(t)$ is  
drawn from some probability distribution $\pi(\eta)$, and is 
independent of it's value at previous time steps.  
This exponential notation is particularly
convenient for working with multiplicative noise, keeping the necessary
algebra at minimum. Under these rules of dynamics the logarithm of
the asset's price, $\ln p(t)$, performs a random walk with an average drift
$v=\avg{\eta}$ and a dispersion $D=\avg{\eta^2}-\avg{\eta}^2$.

It is easy to derive the time evolution of the total capital $W(t)$
of an investor, following the above strategy: 
\begin{equation}
W(t+1)=(1-r) W(t) + r W(t) e^{\eta (t)}
\label{evol}
\end{equation}

Let us assume that the value of the investor's capital at $t=0$ is
$W(0)=1$.  The evolution of the expectation value of the expectation
value of the total capital
$\avg{W(t)}$ after $t$ time steps is obviously given by the recursion
$\avg{W(t+1)}=(1-r+ r\avg{e^{\eta}}) \avg{W(t)}$.  When
$\avg{e^{\eta}}>1$, at first thought the investor should invest all
his money in the risky asset. Then the expectation value of his capital
would enjoy an exponential growth with the fastest growth rate.
However, 
it would be totally unreasonable
to expect that in a {\it typical} realization of price
fluctuations, the investor would be able to attain the {\it average}
growth rate determined as $v_{avg}=d \avg{W(t)}/dt$.  This is because
the main contribution to the expectation value $\avg{W(t)}$ comes from
exponentially unlikely outcomes, when the price of the asset after a
long series of favorable events with $\eta>\avg{\eta}$ becomes
exponentially big. Such outcomes lie well beyond reasonable
fluctuations of $W(t)$, determined by the standard deviation
$\sqrt{Dt}$ of $\ln W(t)$ around its average value 
$\avg{ \ln W(t)}= \avg{\eta} t$. For the investor who deals with just one
realization of the multiplicative process it is better not to rely on
such unlikely events, and maximize his gain in a {\it typical} outcome
of a process. To quantify the intuitively clear concept of a {\it
typical} value of a random variable $x$, we define $x_{typ}$ as a
median\cite{gnedenko} of its distribution, i.e $x_{typ}$ has the property 
that Prob$(x>x_{typ})=$ Prob$(x<x_{typ})=1/2$.  In a multiplicative process
\req{evol} with $r=1$, $W(t+1)=e^{\eta (t)} W(t)$, one can show that
$W_{typ}(t)$ -- the typical value of $W(t)$ -- grows exponentially in
time: $W_{typ} (t)=e^{ \avg{\eta} t}$ at a rate
$v_{typ}=\avg{\eta}$, while the expectation value $\avg{W(t)}$ also
grows exponentially as $\avg{W(t)}=\avg{e^{\eta}}^t$, but at a 
faster rate given by $v_{avg}=\ln \avg{e^{\eta}}$. Notice that $\avg{
\ln W(t)}$ always grows with the {\it typical} growth rate, since
those very rare outcomes when $W(t)$ is exponentially big, do not make
significant contribution to this average.

The question we are going to address is: which investment fraction $r$
provides the investor with the best {\it typical} growth rate
$v_{typ}$ of his capital. Kelly\cite{kelly} has answered this
question for a particular realization of multiplicative 
stochastic process, 
where the capital is multiplied by 2 with probability $q > 1/2$, and
by 0 with probability $p=1-q$.  This case is realized in a gambling
game, where betting on the right outcome pays 2:1, while you know the
right outcome with probability $q>1/2$. In our notation this case
corresponds to $\eta$ being equal to $\ln 2$ with probability $q$ and
$-\infty$ otherwise.  The player's capital in Kelly's model with $r=1$
enjoys the growth of expectation value $\langle W (t) \rangle$ at a rate
$v_{avg}=\ln 2q>0$. In this case it is however particularly clear that
one should not use maximization of the expectation value of the
capital as the optimum
criterion. If the player indeed bets all of his capital at every time
step, sooner or later he will loose everything and would not be able to
continue to play.  In other words, $r=1$ corresponds to the worst {\it
typical} growth of the capital: asymptotically the player will
be bankrupt with probability 1. In this example it is also very
transparent, where the positive {\it average} growth rate comes from:
after $T$ rounds of the game, in a very unlikely (Prob $=q^T$) event
that the capital was multiplied by $2$ at all times (the gambler
guessed right all the time!), the capital is equal to $2^T$.  This 
exponentially large value of the capital outweighs exponentially small
probability of this event, and gives rise to an exponentially growing
average. This would offer condolence to a  gambler who lost everything.

In this chapter we generalize Kelly's arguments for arbitrary 
distribution $\pi(\eta)$.  As we will see this generalization
reveals some hidden results, not realized in Kelly's ``betting''
game. 
As we learned above, the growth of the {\it typical} value of $W(t)$,
is given by the drift of $\avg{\ln W (t)}=v_{typ} t$, which in
our case can be written as
\begin{equation}
v_{typ} (r)=\int d \eta \ \pi (\eta) \ \ln(1+r(e^{\eta}-1))
\label{v_typ}
\end{equation}
One can check that $v_{typ} (0)=0$, since in this case the whole
capital is in the form of cash and does not change in time.  In
another limit one has 
$v_{typ} (1)=\langle \eta \rangle$, since in this case the whole
capital is invested in the asset and enjoys it's typical growth 
rate ($\avg{\eta}=-\infty$ for Kelly's case).  Can one do better
by selecting $0<r<1$? To find the maximum of $v_{typ} (r)$ one
differentiates (\ref{v_typ}) with respect to $r$ and looks for a
solution of the resulting equation: $ 0=v'_{typ}(r)=
\int d \eta \ \pi (\eta) \ (e^{\eta}-1)/(1+r(e^{\eta}-1))$ 
in the interval $0 \leq r \leq 1$.  If such a solution exists, it is
unique since $ v''_{typ}(r)= -\int d \eta \ \pi (\eta) \
(e^{\eta}-1)^2/(1+r(e^{\eta}-1))^2<0$ everywhere.  The values of the
$v'_{typ}(r)$ at 0 and 1 are given by $v'_{typ}(0)=\langle e^{\eta}
\rangle -1$, and $v'_{typ}(1)=1-\langle e^{-\eta} \rangle$.  One has
to consider three possibilities:

(1) $\langle e^{\eta} \rangle<1$. In this case $v_{typ}'(0)<0$.  Since
$v_{typ}''(r)<0$, the maximum of $v_{typ}(r)$ is realized at $r=0$ and
is equal to 0. In other words, one should never invest in an asset
with negative average return per capital $\avg{e^{\eta}}-1<0$.

(2) $\langle e^{\eta} \rangle>1$ , and $\langle e^{-\eta} \rangle>1$.
In this case $v_{typ}'(0)>0$, but $v_{typ}'(1)<0$ and the maximum of
$v(r)$ is realized at some $0<r<1$, which is a unique solution to
$v_{typ}'(r)=0$.  The typical growth rate in this case is always
positive (because you could have always selected $r=0$ to make it 
zero), but not as big as the average rate $\ln \langle e^{\eta} \rangle$,
which serves as an unattainable ideal limit. An intuitive understanding
of why one should select $r<1$ in this case comes from the following
observation: the condition $\langle e^{-\eta} \rangle>1$ makes
$\langle 1/p(t) \rangle$ to grow exponentially in time.  Such an
exponential growth indicates that the outcomes with very small $p(t)$
are feasible and give dominant contribution to
$\avg{1/p(t)}$.  This is an indicator that the asset price is unstable
and one should not trust his whole capital to such a risky
investment.

(3) $\langle e^{\eta} \rangle>1$ , and $\langle e^{-\eta} \rangle<1$.
This is a safe asset and one can invest his whole capital in it.  The
maximum $v_{typ}(r)$ is achieved at $r=1$ and is equal to
$v_{typ}(1)=\ln \langle \eta \rangle$. A simple example of this type
of asset is one in which the price $p(t)$ with equal probabilities is
multiplied by $2$ or by $a=2/3$. As one can see this is a marginal
case in which $\avg{1/p(t)}=$const. For $a<2/3$ one should invest only
a fraction $r<1$ of his capital in the asset, while for $a \geq 2/3$
the whole sum could be trusted to it. The specialty of the case with
$a=2/3$ cannot not be guessed by just looking at the typical and
average growth rates of the asset!  One has to go and calculate
$\langle e^{-\eta} \rangle $ to check if $\langle 1/p(t) \rangle$
diverges. This ``reliable'' type of asset is a new feature of the 
model with a general $\pi(\eta)$. It is never realized in Kelly's
original model, which always has $\langle \eta \rangle = -\infty$, 
so that it never makes sense to gamble the whole capital every 
time.

An interesting and somewhat counterintuitive consequence 
of the above results is that under certain conditions one
can make his capital grow by investing in asset with a negative
typical growth rate $\langle \eta \rangle<0$. Such asset certainly
loses value, and its {\it typical} price experiences an exponential
decay. Any investor bold enough to trust his whole capital in such an
asset is losing money with the same rate. 
But as long as
the fluctuations are strong enough to maintain a positive {\it
average} return per capital
$\avg{e^{\eta}}-1>0$) one can maintain a certain fraction 
of his total capital invested in this asset and 
almost certainly make money!   
A simple example of such mind-boggling situation
is given by a random multiplicative process in which the price of the
asset with equal probabilities is doubled (goes up by $100\%$) or
divided by 3 (goes down by $66.7\%$). The typical price of this asset
drifts down by 18\% each time step. Indeed, after $T$ time steps one
could reasonably expect the price of this asset to be
$p_{typ}(T)=2^{T/2}3^{-T/2}=(\sqrt{2/3})^T \simeq 0.82^T$.  On the other
hand, the average $\langle p(t) \rangle$ enjoys a 17\% 
growth $\langle p(t+1) \rangle=7/6 \ \langle p(t) \rangle \simeq 1.17\langle W(t)\rangle$. As one can easily see, the optimum of the typical growth rate is
achieved by maintaining a fraction $r=1/4$ of the capital invested in
this asset.  The typical rate in this case is a meager $\sqrt{25/24}
\simeq 1.02$, meaning that in a long run one almost certainly 
gets a 2\% 
return per time step, but it is certainly better than losing 18\% 
by investing the whole capital in this asset. 

The temporal evolution of another example
is shown in the Figure 1, where a risky asset varies daily by 
+30\% 
or -24.4\% 
with equal chance,
this is not unlike $daily$ variation of some 
"red chips" quoted in Hong Kong or some Russian 
companies quoted on the Moscow Stock Exchange. 
In this example, the stock is almost certainly doomed: 
in the realization shown on Fig. 1 in four years 
the price of one share went down by
a factor $500$, it was practically wiped out.
At the same time the investor maintaining the optimal $\simeq 38 \%$ 
investment fraction profited handsomely, making 
more than 500 times of his 
starting capital! It is all the more remarkable that 
this profit is achieved without any insider 
information but only by dynamically
managing his investment in such a bad stock.

Of course the properties of a typical realization of a random
multiplicative process are not fully characterized by the drift
$v_{typ} (r) t$ in the position of the center of mass of $P(h, t)$,
where $h(t)=\ln W(t)$ is a logarithm of the wealth of the investor.
Indeed, asymptotically $P(h,t)$ has a Gaussian shape $P(h,t)={1 \over
\sqrt{2 \pi D(r) t}}
\exp (- {(h-v_{typ} (r) t)^2 \over 2 D (r) t})$, where
$v_{typ}(r)$ is given by eq. \req{v_typ}.  One needs to know the
dispersion $D(r)$ to estimate $\sqrt{ D(r) t}$, which is the
magnitude of characteristic deviations of $h(t)$ away from its typical
value $h_{typ} (t) = v_{typ} t$.  At the infinite time horizon $t \to
\infty$, the process with the biggest $v_{typ}(r)$ will certainly be
preferable over any other process.  This is because the separation
between typical values of $h(t)$ for two different investment
fractions $r$ grows linearly in time, while the span of typical
fluctuations grows only as a $\sqrt{t}$.  However, at a finite time
horizon the investor should take into account both $v_{typ}(r)$ and
$D(r)$ and decide what he prefers: moderate growth with small
fluctuations or faster growth with still bigger fluctuations. To
quantify this decision one needs to introduce an investor's
``utility function'' which we will not attempt in this work.  The most
conservative players are advised to always keep their capital in cash,
since with any other arrangement the fluctuations will certainly be
bigger.  As a rule one can show that the dispersion $D(r)=\int
\pi(\eta) \ln^2[1+r(e^{\eta}-1)] d \eta - v_{typ}^2$ monotonically
increases with $r$. Therefore, among two solutions with equal
$v_{typ}(r)$ one should always select the one with a smaller $r$,
since it would guarantee smaller fluctuations.

We proceed with deriving analytic results for the optimal
investment fraction $r$ in a situation when fluctuations of asset
price during one readjustment period (one step of the discrete 
dynamics) are small. This
approximation is usually justified for developed markets, if the
investor sells and buys asset to maintain his optimal ratio on let's
say monthly basis.  Indeed, the month to month fluctuations in, for
example, Dow-Jones Industrial Average 
i) to a good approximation are uncorrelated random numbers; 
ii) seldom raise above few percent, 
so that the assumption that $\eta (t) \ll 1$ is justified.

Here it is more convenient to switch to the standard 
notation. It is customary to use the random variable
\beq
\Lambda(t)={p(t+1)-p(t) \over p(t)}=e^{\eta (t)}-1, 
\eeq 
which is referred to as return per unit capital of the asset.  The
properties of a random multiplicative process are expressed in terms
of the average return per capital
$\alpha=\avg{\Lambda}=\avg{e^{\eta}}-1$, and the volatility (standard
deviation) of the return per capital
$\sigma=\sqrt{\avg{\Lambda^2}-\avg{\Lambda}^2}$.  In our notation
$\alpha=\avg{e^{\eta}}-1$ is determined by the {\it average} and not
typical growth rate of the process.  For $\eta \ll 1$ , $\alpha \simeq
v+D/2+v^2/2$, while the volatility $\sigma $ is related to $D$ (
the dispersion of $\eta$) through $\sigma \simeq \sqrt{D}$.

Expanding Eq. \req{v_typ} up to the second order in
$\Lambda=e^{\eta}-1$ one gets: $v_{typ} \simeq
\avg{r(e^{\eta}-1)-r^2(e^{\eta}-1)^2}=
\alpha r-(\sigma^2+\alpha^2)r^2/2$.
The optimal $r$ is given by
\beq
r_{opt}={\alpha \over \alpha^2+\sigma^2}
\label{r_opt}
\eeq 
If the above formula prescribes $r_{opt}>1$, the investor is
advised to trust his whole capital to this asset.  We remind you
that in this paper the risk-free return per capital is set to zero 
(investor keeps the rest of his capital in cash).
In a more realistic case, when a risk-free
bank deposit brings a return $p$ during a single
readjustment interval, the formula for the optimal investment 
ratio should be generalized to: 
\beq
r_{opt}={\alpha - p \over \alpha^2+\sigma^2}.
\eeq

In a hypothetical case discussed by Merton\cite{merton},  
when asset's price follows a continuous
multiplicative random walk (i.e. price fluctuations are uncorrelated
at the smallest time scale) and the investor is committed to 
adjust his investment ratio on a continuous basis, one should 
use infinitesimal quantities $\alpha \to \alpha dt$ and
$\sigma^2 \to \sigma^2 dt$. Under these 
circumstances the term $\alpha^2 dt^2$, being second order in 
infinitesimal time increment $dt$,  should be dropped
from the denominator. Then one recovers an optimal investment fraction
for ``logarithmic utility'' derived by Merton\cite{merton}. 

Asset price fluctuations encountered in developed financial markets have
relatively large average returns and small volatilities, so that the
optimal investment fraction into any given asset $r_i^{opt}$ is almost
always bigger than $1$. For instance the data for average annual
return and volatility of Dow-Jones index in 1954-1963\cite{sharpe}
are $\alpha_{DJ}=16\%$, $\sigma_{DJ}=20\%$, while the average
risk-free interest rate is $p=3\%$. This suggests that for 
an investor committed to yearly readjustment of his asset holdings 
to the selected ratio, the optimal
investment ratio in Dow-Jones portfolio is
$r_{DJ}=(\alpha_{DJ}-p)/(\sigma_{DJ}^2+\alpha_{DJ}^2)=1.98>1$.  
On the other hand the investor ready to readjust his stock holdings
every month should use $\alpha_{\rm monthly} \simeq \alpha/12$ and 
$\sigma_{\rm monthly} \simeq \sigma/\sqrt{12}$. For him the optimal
investment fraction would be $r_{DJ}^{\rm
monthly}=(\alpha_{DJ}/12-p/12)/(\sigma_{DJ}^2/12+(\alpha_{DJ}/12)^2)
\simeq 3.09$.
In both cases, given no other alternatives the investor interested in a
long-term capital growth is advised to trust his whole capital
to Dow-Jones portfolio and enjoy a {\it typical} annual return
$\alpha-\sigma^2/2=14\%$, which is $2\%$ smaller than the average
annual return of $16\%$ but significantly bigger than the 
risk-free return of 3\%.

\section{Optimization of multi-asset portfolio}
\noindent
We proceed by generalizing the results of a previous chapter to a more
realistic situation, where the investor can keep a fraction of his total
capital in a portfolio composed of $N$ risky assets. 
The returns per unit capital of different assets 
are defined as $\Lambda_i(t)={p_i(t+1)-p_i(t) \over
p_i(t)}=e^{\eta_i}-1$.  
Each asset is characterized by an average return per capital
$\alpha_i=\avg{e^{\eta_i}}-1$, and volatility
$\sigma_i=\sqrt{\avg{e^{2 \eta_i}}-\avg{e^{\eta _i}}^2}$.  As in the
single asset case, an investor has decided to maintain a given fraction
$r_i$ of his capital invested in $i$-th asset, and to keep the rest in
cash. His goal is to maximize the {\it typical} growth rate of his
capital by selecting the optimal set of $r_i$. The explicit expression
for the typical rate under those circumstances is given by
\beq
v_{typ}(r_1, r_2 \ldots r_N)=\avg{\ln[1+\sum_{i=1}^N
r_i(e^{\eta_i}-1)]}.
\label{typ_N}
\eeq
The task of finding an analytical solution for the global maximum of
this expression seems hopeless.  We can, however, expand the logarithm
in eq. \req{typ_N}, assuming that all returns $\Lambda
_i=e^{\eta_i}-1$ are small. Then to a second order one gets:
$v_{typ}=\sum_i \alpha_i r_i - \sum_{i,j} K_{ij} r_i r_j/2$, where
$K_{ij}$ is a covariance matrix of returns, defined by
$K_{ij}=\avg{\Lambda_i \Lambda_j}$.  In this work we restrict
ourselves to the case of uncorrelated assets, when the only nonzero
elements of covariance matrix lay on the diagonal,
$K_{ij}=(\alpha_i^2+\sigma_i^2) \delta_{ij}$.  
In this case the expression for typical rate becomes
\beq
v_{typ}=\sum_{i=1}^N [\alpha_i r_i - (\sigma_i^2 +\alpha_i^2)r_i^2/2],
\label{typ_N2}
\eeq
without any restrictions 
the optimal investment fraction in a given asset is determined by
a single asset formula \req{r_opt}
\beq
\tilde{r_i}^{opt}= {\alpha_i \over \sigma_i ^2+ \alpha_i^2}
\label{r_opt2}
\eeq
In case of the general covariance matrix the above formula 
becomes
\beq
\tilde{r_i}^{opt}= \sum _j (K^{-1})_{ij} \alpha_j, 
\eeq
where $(K^{-1})_{ij}$ is an element of a matrix inverse to $K_{ij}$.
With somewhat heavier algebra all results from the following
paragraphs can be reformulated to include the effects of a general
covariance matrix and non-zero risk-free interest rate.
However, we will not attempt it in this manuscript.

The nontrivial part of the $N$ asset case comes from the restriction
$\sum r_i \leq 1$. This restriction starts to be relevant if $\sum_i
\tilde{r_i}^{opt}>1$, and the Eq. \req{r_opt2} no longer 
works. In this case the optimal solution would be to
invest the whole capital in assets and to search for a maximum of
$v_{typ}$ restricted to the hyperplane $\sum_i r_i=1$. Unfortunately, this
interesting case was overlooked by Merton\cite{merton}.
Therefore his prescription for the vector of optimal investment
fractions holds only in quite unrealistic situation when 
$\sum \alpha_i/\sigma_i^2 \leq 1$.
Introducing a Lagrange
multiplier $\lambda$, one gets
$r_i^{opt}=(\alpha_i-\lambda)/(\alpha_i^2+\sigma_i^2)$.  Obviously, the
assets for which $r_i<0$ should be dropped and the optimal $r_i^{opt}$
are finally given by 
\beq
r_i^{opt}={\alpha_i-\lambda \over
\alpha_i^2+\sigma_i^2} \theta({\alpha_i-\lambda \over 
\alpha_i^2+\sigma_i^2}),  
\label{r_opt3}
\eeq
where $\theta(x)$ is a usual Heavyside step function.  The Lagrange
multiplier $\lambda$ is found by solving
\beq
\sum_{i=1}^N {\alpha_i-\lambda \over 
\alpha_i^2+\sigma_i^2} \theta ({\alpha_i-\lambda \over
\alpha_i^2+\sigma_i^2})=1
\label{lagr}
\eeq

To demonstrate how this optimization works in practice we consider the
following simple example. An investor has an alternative to invest his
capital in 3 assets with average returns $\alpha_1=1.5\%$,
$\alpha_2=2\%$, $\alpha_3=2.5\%$. Each of these assets has the 
same volatility $\sigma=10\%$.
Which are optimal investment
fractions in this case?  The eq. \req{r_opt2} recommends
$\tilde{r_1}^{opt}=\alpha_1/(\sigma^2+\alpha_1^2) \simeq 
\alpha_1/\sigma^2=1.5$, $\tilde{r_2}^{opt} \simeq 2$, 
$\tilde{r_3}^{opt} \simeq 2.5$. Each of
these numbers is bigger than one, which means that given any one of
these assets as the only investment alternative, the investor would be
advised to trust his whole capital to it. As was explained above,
whenever the eq. \req{r_opt2} results in $\tilde{r_1}^{opt}+
\tilde{r_2}^{opt}+\tilde{r_3}^{opt}>1$, the investor
should not keep any money in cash. We need to solve the
eq. \req{r_opt3} to determine how he should share his capital between
three available assets. Assuming first that each asset gets a nonzero
fraction of the capital, one writes the equation \req{lagr} for the
Lagrange multiplier $\lambda$: $1.5-\lambda+2-\lambda+2.5-\lambda=1$,
or $\lambda=5/3 \simeq 1.67$.  But then $r_1=1.5-\lambda$ is
negative. This suggests that the average return in asset 1 is too
small, and that the whole capital should be divided between assets 2
and 3. Then the eq. \req{lagr} $2-\lambda+2.5-\lambda=1$ has the
solution $\lambda=1.75$, and the optimal investment fractions are
$r_1^{opt}=0$, $r_2^{opt}=0.25$, $r_3^{opt}=0.75$.  This optimum
represents the compromise between the following two tendencies.  On
one hand, diversification of the portfolio tends to 
increase its typical growth rate and bring it closer to 
the average growth rate. This happens because fluctuations
of different asset's prices partially cancel each other 
making the whole portfolio less risky. But, on
the other hand, to diversify the portfolio one has 
to use assets with $\alpha$'s smaller than that of the
best asset in the group, and thus compromise the average growth rate
itself. In the above example the average return $\alpha_1$ was just
too low to justify including it in the portfolio.

Finally, we want to compare our results with the exact formula derived by 
Breiman\cite{breiman}. His argument goes as follows: in case where 
there is no bank (or it is just included as the alternative of investing in
a risk-free asset for which $\Lambda=p$ and $\sigma=0$) one wants to
maximize $\avg{\ln \sum r_i e^{\eta_i}}$ subject to the 
constraint $\sum r_i=1$. Introducing a Lagrange multiplier $\beta$
(different from Lagrange multiplier $\lambda$ used above) one gets
a condition for an extremal value of growth rate:
$\avg{e^{\eta_i}/\sum r_i e^{\eta_i}}-\beta=0$. This can be also
written as $\avg{r_i e^{\eta_i}/\sum r_i e^{\eta_i}}-\beta r_i=0$.
The summation over $i$ shows that $\beta=1$, therefore at optimum
is determined by a solution of the system of $N$ equations:
\beq
r_i=\avg{r_i e^{\eta_i}/\sum r_i e^{\eta_i}}. 
\label{breiman_eq}
\eeq 
notice that the $i$th equation automatically holds if $r_i=0$. 
Therefore, finding an optimal set of investment fractions $r_i$ 
is equivalent to solving \req{breiman_eq} with $r_i \geq 0$.
According to this equation in the strategy, optimal in Kelly's sense, 
{\it on average} one does not have to buy or sell assets since the
{\it average}
fraction of each asset's capital in the total capital
( $\avg{r_i e^{\eta_i}/\sum r_i e^{\eta_i}}$ ) is conserved by dynamics.
Unfortunately, the exact set of equations \req{breiman_eq} 
is as unusable as it is elegant: it suggests
no constructive way to derive the set of optimal investment 
fractions from known asset's  average returns and covariance matrix. 
In this sense our set of approximate 
equations \req{r_opt3} provides an investor with a constructive method
to iteratively determine the set of optimal weights of different
assets in the optimal portfolio.

The work at Brookhaven National Laboratory was 
supported by the U.S. Department of Energy Division
of Material Science, under contract DE-AC02-76CH00016.
S.M. and thanks the Institut de Physique Th\'eorique 
for the hospitality during the visit, when this work was 
initiated. 
This work is supported in part by the Swiss National
Foundation through the Grant 20-46918.96.

\nonumsection{References}

\begin{figure}[htbp]
\vspace*{13pt}
\centerline{\vbox{\hrule width 5cm height0.001pt}}
\vspace*{1.4truein}		
\centerline{\vbox{\hrule width 5cm height0.001pt}}
\vspace*{13pt}
\fcaption{
Temporal evolution of the stock and the optimizing investor's capital. 
The time units can be interpreted as days and the total period (1000 days) 
is about 4 years. During this period the doomed stock performed very
badly, whereas our investor
made huge profit from investing in it dynamically with $r \simeq 38\%$. 
Not only the optimal strategy performs better,
it also has much less volatility.}
\end{figure}

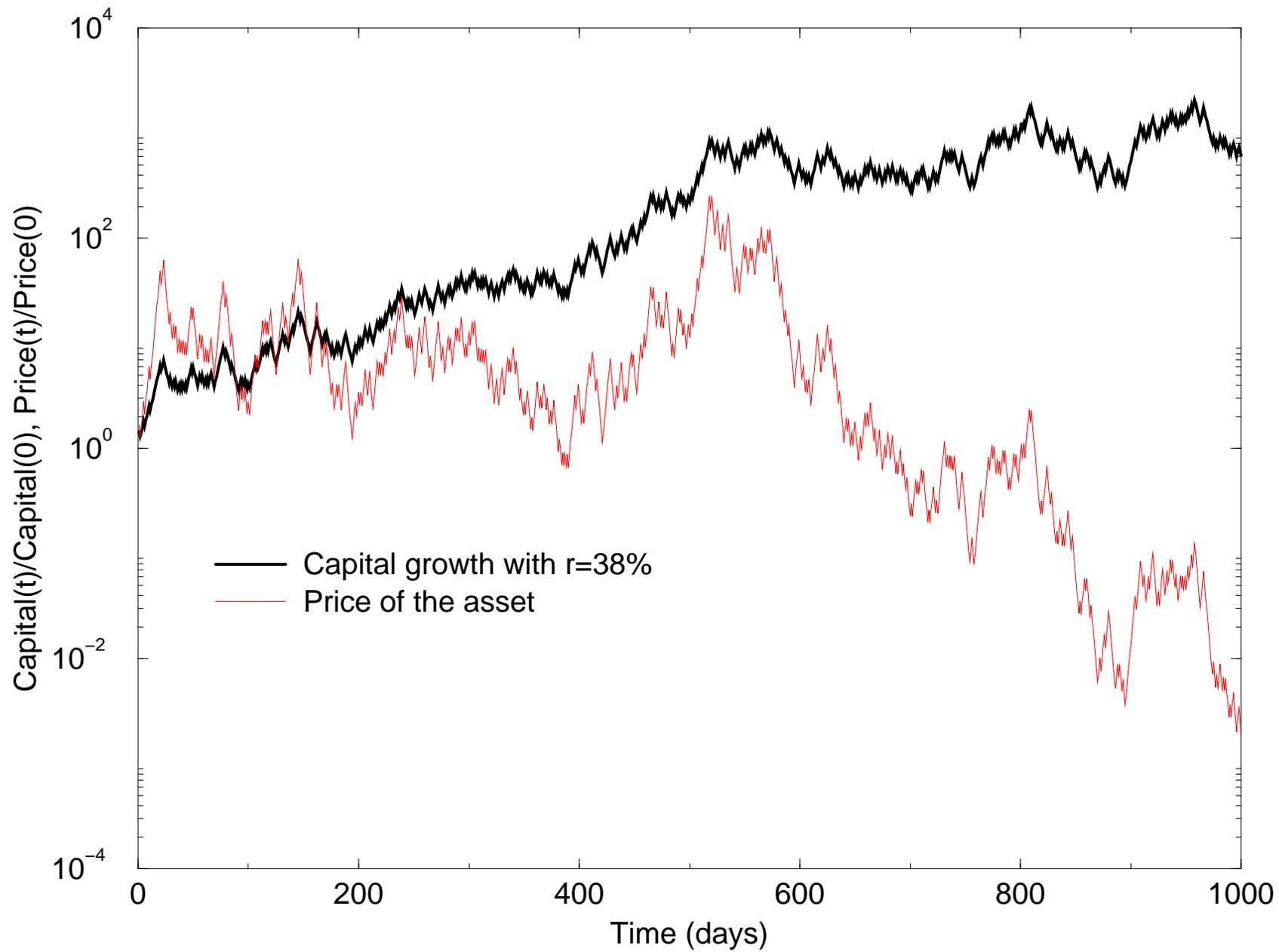
\end{document}